\documentclass[11pt,letterpaper]{article}

\usepackage{chemformula} 
\usepackage[T1]{fontenc} 
\usepackage{gensymb}
\usepackage{mciteplus}
\usepackage{float}
\usepackage{amsmath}
\usepackage{graphicx}
\usepackage{geometry}
\let\oldAA\AA
\renewcommand{\AA}{\text{\normalfont\oldAA}}

\begin{document}

\begin{huge}
\noindent Extended MARTINI Water Model for Heat Transfer Studies
\end{huge}

\textbf{Sumith Yesudasan}

sumith.yesudasan@uj.edu

Department of Mechanical Engineering, University of Jamestown, Jamestown, ND

\section*{Abstract}
The computationally efficient classical MARTINI model is extended to simulate heat transfer simulations of water. The current MARTINI model, variations of it and other coarse grain water models focus on reproducing the thermodynamic properties below room temperature, hence making them unsuitable for studying high temperature simulations especially evaporation at 100 \degree C. In this work, the MARTINI model is reparametrized using a combination of Genetic Algorithm, Artificial Neural Network and Nelder-Mead optimization technique to match the phase equilibrium properties of water. The reparametrized model (MARTINI-E) accurately reproduces density, enthalpy of vaporization and surface tension at 100\degree C and outperforms other leading coarse grain water models. The model is also validated using the energy conservation and enthalpy change due to latent heat in a lamellar system. This new water model can be used for simulating phase change phenomena, thin film evaporation and other energy transport mechanisms accurately.

\textbf{Keywords}: CGMD, water models, Molecular Dynamics, MARTINI, Optimization
\section{Introduction}

In the semi-conductor industry and high-concentration photovoltaics, the emergence of higher energy density calls for highly efficient and faster cooling systems. Out of many conventional cooling techniques, evaporative water cooling stands out as a best candidate due to its high specific heat capacity and high enthalpy of vaporization. Based on kinetic theory \cite{gambill1989upper}, an evaporative heat flux of $20,000 W/cm^2$ can be achieved using water. However, even with the recent developments in Nano-scale fabrication techniques the maximum heat flux is limited to 500 to 1000 $W/cm^2$ \cite{jaikumar2017microscale}. This points out to our poor understanding of the nanoscale and microscale evaporation of water. Studies show that thin film evaporation has high potential to remove heat compared to bulk region \cite{bigham2017physics,wang2018ultrahigh}. Studying water evaporation at nanoscale using experiments is a challenging attempt. Use of molecular dynamics can become helpful in this scenario \cite{akkus2019molecular,akkus2019atomic, yd2015surface}, however there is no single water model which can capture or simulate all of its properties \cite{jorgensen1983comparison}. Even the best performing and widely used models like SPCE \cite{berendsen1987missing} and TI3P \cite{jorgensen1983comparison} are unsuitable to study evaporation, due to the computational cost. This shifts our focus to computationally faster models called coarse grain molecular dynamics (CGMD) models.

In a typical CGMD model, one or more water molecules are combined into a bead or a super-molecule to represent the bulk properties of the system. Most of the existing CGMD models are developed for bio-molecular studies \cite{marrink2007martini,murtola2004coarse,vishnyakov2018coarse, yesudasan2018molecular, yesudasan2017coarse, yesudasan2018coarse, yesudasan2018fibrin, yesudasan2018multiscale} and is not tested for heat transfer. The existing CGMD models \cite{vishnyakov2018coarse,raubenolt2018coarse} mainly focus on the room temperature behavior of water or even sometimes the behavior below zero degree Celsius. While these models can capture freezing, ice formation and other properties of water \cite{chan2019machine}, their applicability to high temperature applications is limited or seldom. The field of coarse graining itself is a big research area and hence the readers are advised to refer consolidated reviews found elsewhere \cite{hadley2012coarse, riniker2012developing, brini2013systematic}. Among various types of CGMD models, mono bead models are appealing due to its very low computational power consumption.

In this work, an extension to the existing CGMD model called MARTINI \cite{marrink2004coarse, marrink2007martini} is investigated. The mono beads represents four water molecules and interact with each other using a force field called soft Lennard-Jones (LJ) potential \cite{beutler1994avoiding} . Using the evolutionary algorithm called genetic algorithm, the initial population (set of force field parameters) is optimized to get a local cost minima. The cost is estimated by comparing the CGMD simulation results with experimental values of density, surface tension and enthalpy of vaporization. A simple network of artificial neural network (ANN) is used to predict the next best population during the iteration of genetic algorithm. Local cost minimum is achieved by applying Nelder-Mead optimization to the converged solution of genetic algorithm. Such obtained parameters are able to simulate surface tension, density at constant temperature (100 \degree C) and pressure and enthalpy of vaporization of water with less than 1\% error. This new optimized model called as MARTINI-E (MARTINI Extended) can simulate evaporation of water, heat transfer with solids, boiling and other important thermodynamics \cite{yd2016origin, yesudasan2016molecular} at nanoscale using CGMD simulation.

\section{Coarse Grain Models of water}

\begin{figure}[H]
\includegraphics[width=\textwidth]{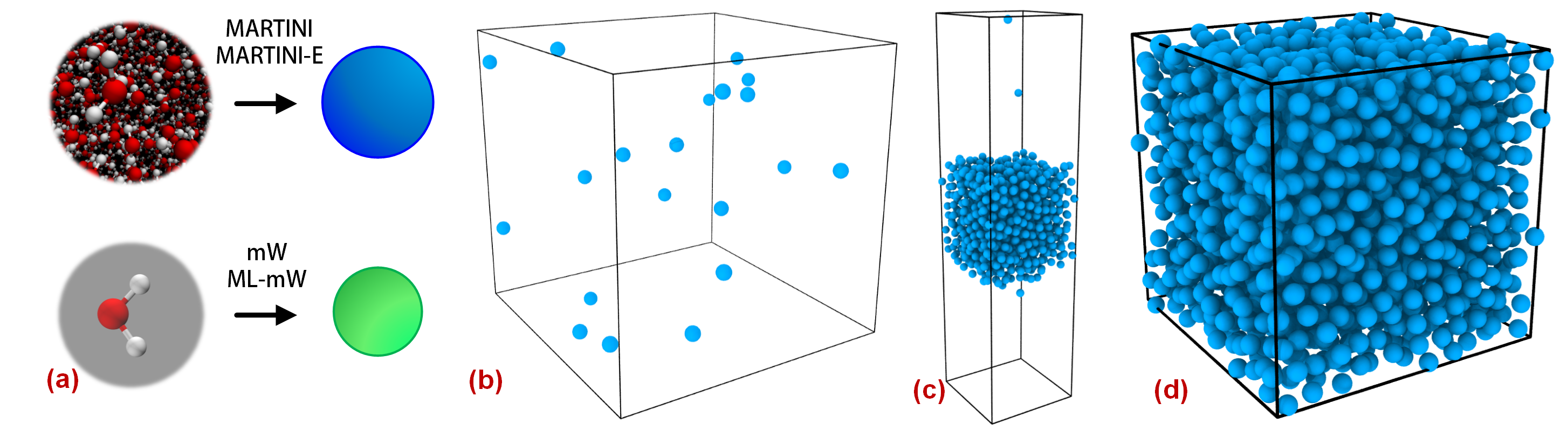}
  \caption{(a) Formulation of different types of CGMD models, (b) Representative vapor CGMD system, (c) Vapor-Liquid-Vapor CGMD system, (d) Liquid CGMD system}
  \label{fig1}
\end{figure}

There exists a wide range of coarse grain water models in the literature, extending from single water model to analytical multi water models \cite{hadley2012coarse}. The present work focuses only on the CGMD models with a single bead. Figure \ref{fig1}a shows the coarse graining of water molecules into a bead. Figure \ref{fig1}b-d shows CGMD models used for simulating vapor, film and liquid phases of water.

\subsection{Existing Coarse Grain Models}

Among them, the classical MARTINI water model \cite{marrink2004coarse, marrink2007martini}, mW model \cite{molinero2008water}, and the recently developed ML-mW model \cite{chan2019machine}  can simulate most of the thermodynamic properties at temperatures near and below 300 K. However, their feasibility near 373.15 K (100 \degree C) is not tested.

\subsubsection{MARTINI model}
The MARTINI CGMD model maps a cluster of four water molecules to a single bead \cite{marrink2004coarse,marrink2007martini} and interacts with each other using a standard 12-6 Lennard Jones (LJ) potential (Eq. \ref{equation1}). 

\begin{equation}
  E_{MARTINI} = 4 \epsilon \left[ \left( \frac{\sigma}{r} \right)^{12} -   \left( \frac{\sigma}{r}  \right)^6  \right]
  \label{equation1}
\end{equation}

Here, for water, $\epsilon = 5 ~kJ/mol=1.19503 ~kcal/mol$ and $\sigma=4.7 ~\AA$ are standard parameters. In the computer simulations this LJ interaction potential is shifted (both potential and force smoothly goes to zero) and a cutoff distance $r_{cut}=12 ~\AA$, corresponding to approximately $2.5\sigma$ is used. These parameters and potential function corresponds to MARTINI 2.0 \cite{marrink2007martini}. While MARTINI models are simple in construction, they need additional anti-freeze particles to keep water from freezing at temperatures as high as 300 K and also have poor representation of compressibility and surface tension \cite{hadley2012coarse}.

\subsubsection{Stillinger-Weber based models}

In the mono bead category, the next leading CMGD model is called mW model \cite{molinero2008water} which is based on Stillinger-Weber (SW) potential \cite{stillinger1985computer}. In mW model, one water molecule is mapped onto one coarse grain bead and water is modeled as an intermediate element between carbon and silicon. The functional form of SW potential is as shown in below equations (\ref{equation2}, \ref{equation3}, \ref{equation4}).

\begin{equation}
E=\sum_i \sum_{j>i} \phi_2(r_{ij}) \sum_i \sum_{j\neq i}\sum_{k>j} \phi_3(r_{ij},r_{ik},\theta_{ijk})
  \label{equation2}
\end{equation}
\begin{equation}
\phi_2 (r_{ij} )=A_{ij} \epsilon_{ij} \left[B_{ij} \left(\frac{\sigma_{ij}}{r_{ij}} \right)^{p_{ij}}- \left(\frac{\sigma_{ij}}{r_{ij}} \right)^{q_{ij}} \right] \exp{\left( \frac{ \sigma_{ij}}{r_{ij} - a_{ij}\sigma_{ij}} \right)}
  \label{equation3}
\end{equation}
\begin{equation}
    \phi_3 (r_{ij}, r_{ik}, \theta_{ijk}) = \lambda_{ijk} \epsilon_{ijk} \left[\cos{\theta_{ijk}} - \cos{\theta_{0ijk}} \right]^2  \exp{\left( \frac{\gamma_{ij} \sigma_{ij}}{r_{ij} - a_{ij}\sigma_{ij}} \right)} \exp{\left( \frac{\gamma_{ik} \sigma_{ik}}{r_{ik} - a_{ik}\sigma_{ik}} \right)}
    \label{equation4}
\end{equation}

 The parameters of mW model are further improvised recently using machine learning algorithms \cite{chan2019machine} and is called as ML-mW model. The description of the variables of SW potential and the parameters for mW and ML-mW are described elsewhere in literature \cite{molinero2008water, chan2019machine, stillinger1985computer}. Even though mW and ML-mW can simulate thermodynamic properties of water better than MARTINI model, the computational cost is higher due to the smaller number of water molecules mapping per bead and also the presence of three body potential functions. Moreover, the latter two models mainly focus on solidification of water and thermodynamics around 0 \degree C.

\subsection{Extending the MARTINI model}

The MARTINI model can represent 4 water molecules and can have larger timestep of integration (20 fs to 30 fs \cite{winger2009using,marrink2010comment}). This makes it a suitable model to investigate for its validity near 373.15 K. Instead of conventional 12-6 LJ potential, a modified form called called soft Lennard-Jones (LJ) potential \cite{beutler1994avoiding} will be used, as shown below.

\begin{equation}
    E = \lambda^n 4 \epsilon \left\{ \left[ \alpha_{LJ} (1 - \lambda)^2 + \left( \frac{r}{\sigma}\right)^6 \right]^{-2} - \left[ \alpha_{LJ} (1 - \lambda)^2 + \left( \frac{r}{\sigma}\right)^6 \right]^{-1} \right\}
    \label{equation5}    
\end{equation}

Here, $\alpha_{LJ}$ is taken as 0.5 and $n$ as 2. The remaining parameters $\epsilon$, $\lambda$ and $\sigma$ are tunable and will be used to parametrize in this study. A simple LJ potential alone will not be sufficient to simulate the viscous nature of the water, hence a viscous force with a parameter ($\gamma$) scaling velocity of the beads, is added to the force field as shown below. 

\begin{equation}
    F = - \gamma ~ v
    \label{equation7}
\end{equation}

\subsubsection{Genetic algorithm phase}
\begin{figure}[H]
\includegraphics[width=\textwidth]{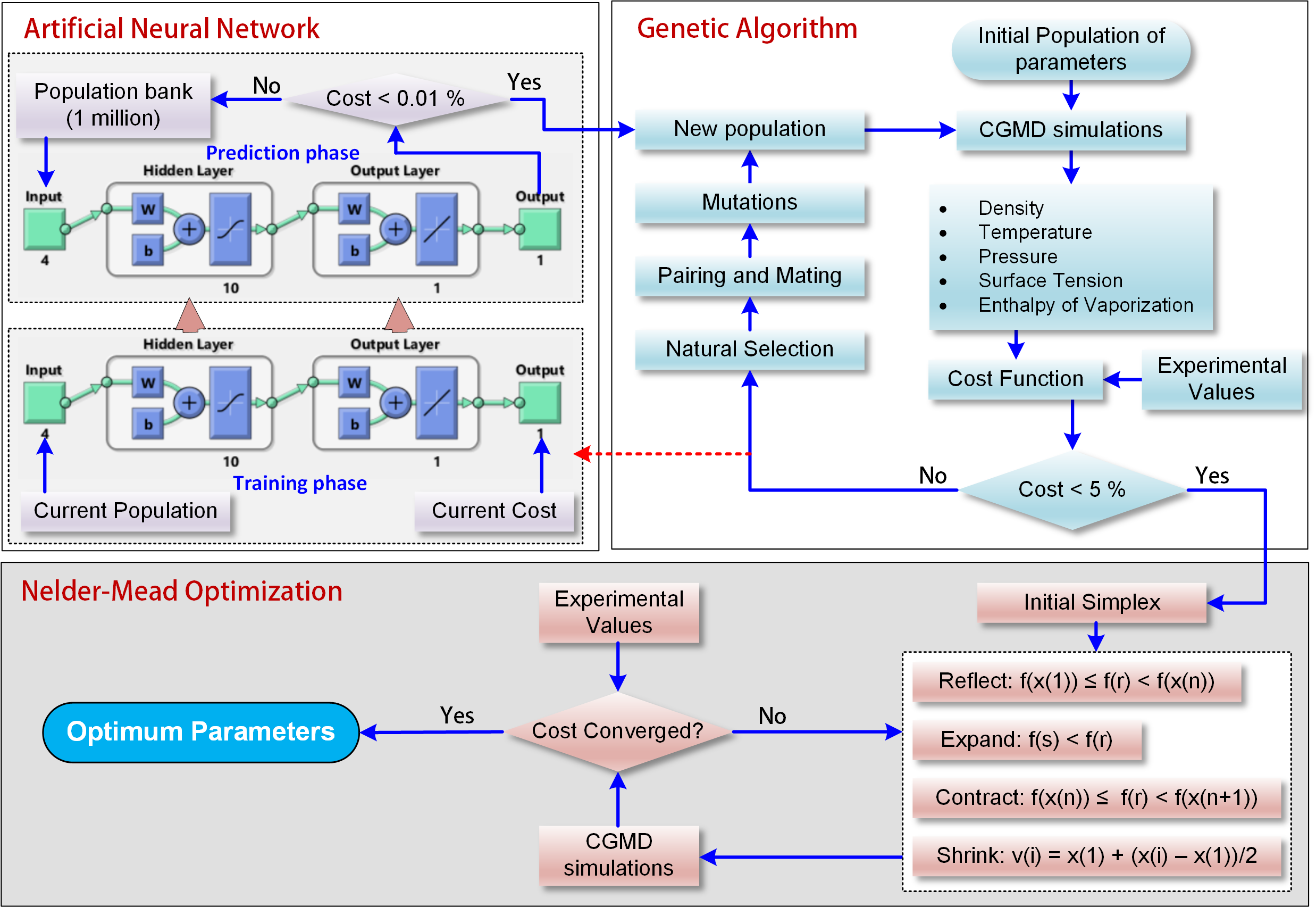}
  \caption{Coarse graining protocol for extended MARTINI model development}
  \label{fig2}
\end{figure}

The coarse graining protocol used in this work is shown in the Fig. \ref{fig2}. It starts with a population of 50, and corresponding parameters for the model and fed into the genetic algorithm. The selection, pairing, mating and mutations are performed based on the textbook by Haupt and Ellen \cite{haupt2004practical}. The selection rate is 50\% and mutation rate is 20\%. The mating step follows the BLX-$\alpha$ method \cite{eshelman1993real}. To create the initial population, a uniform distribution noise within interval $(0,1)$ is created and added to the original MARTINI parameters.

For every population in genetic algorithm, CGMD equilibrium simulations of vapor system, film system and bulk liquid system as shown in Fig. \ref{fig1}b-d  is performed to estimate enthalpy of vaporization ($dh = h_{vapor} - h_{liquid}$), surface tension  and density respectively. The vapor system ($20 nm \times 20 nm \times 20 nm$) consists of 20 beads, film system ($5 nm \times 5 nm \times 20 nm$) consists of 1045 beads and liquid system ($7.3452 nm \times 7.3452 nm \times 7.3452 nm‬$) consists of 2000 beads. Vapor and film systems are equilibrated at 373.15 K using original MARTINI model parameters and NVT ensemble. The liquid system is equilibrated at 373.15 K and $1.00090 ~atm$ (saturation pressure) using NPT ensemble.

All CGMD simulations are performed using LAMMPS software \cite{plimpton1995fast} and the temperature and pressure is controlled using a Nose-Hoover thermostat and barostat \cite{martyna1996explicit,martyna1994constant,shinoda2004rapid}  with a chain length of four. The simulations start with an energy minimization for 1000 steps, followed by equilibration runs of 50000 steps and production runs of 50000 steps. The time step of integration is chosen as $5 fs$ which is relatively low compared to a standard CGMD simulation. Number of steps for equilibration and production are selected based on the time required for the equilibration of energy, temperature and pressure and the standard deviations of energy and temperature is within 1\%. The cut off radius for the simulations are taken as $12 \AA$, same as the original MARTINI model.

From the simulations, diffusion coefficient, density, surface tension and enthalpy of vaporization of the systems are estimated and compared to the experimental values from the NIST standard web book \cite{lemmon2013nist}. This comparison is made using a percentage error based cost function ($C$) as described in the below equation. 

\begin{equation}
    C = \frac{100}{N} \sum_{i=1}^N \frac{f_{EXP}^i - f_{MD}^i}{f_{EXP}^i}
    \label{equation6}
\end{equation}

\begin{figure}[H]
\begin{center}
\includegraphics[width=0.6\textwidth]{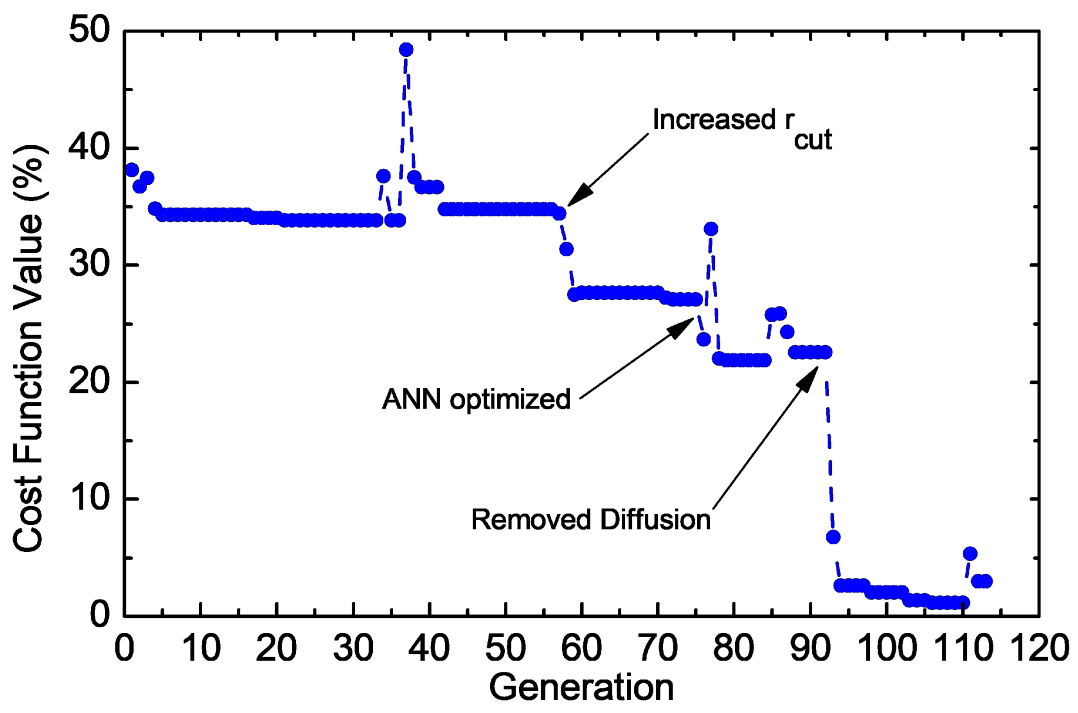}
  \caption{Cost function value over generations of genetic algorithm}
  \label{fig3}
\end{center}  
\end{figure}

Here, $f$ represents the quantities density, surface tension and enthalpy, $N$ is the number of parameters, which is 3 in our case. A cost value closer to 0 means a very good agreement between the model and the experiments. If the score is not close enough to 0, then a new population of parameters is chosen as per the logic given in Fig. \ref{fig2}. The results show that (Fig. \ref{fig3}) even after 50 generations, the cost value was not converging and was fluctuating near 35\%.

\subsubsection{ANN implementation}

At this point, the cut off radius of the system is increased to $18 \AA$. This change has improved the cost function value to lower to upper twenties, but still not converging. So an Artificial Neural Network is introduced to predict the best parameters. The algorithm flowchart is shown in the Fig. \ref{fig2}, where the current population and cost values are fed into a neural network with 10 hidden layers and an output layer of matlab nnstart tool. Using Levenberg–Marquardt algorithm, the network is trained with a R value >0.95. Then a set of 1 million randomly generated population bank is generated with uniform distribution noise around the current best population. This bank is used to search for best parameters which have cost value less than 0.001 \%. Thus selected parameters are used to replace 50\% of the worst population in the original set. Still the cost function value was fluctuating near to 20\%. After careful observation of the individual cost values, we decided to drop the diffusion coefficient from the cost function. This led to a quick convergence (less than 5\%) of the system as shown in the Fig. \ref{fig3}.

\begin{figure}[H]
\begin{center}
\includegraphics[width=0.6\textwidth]{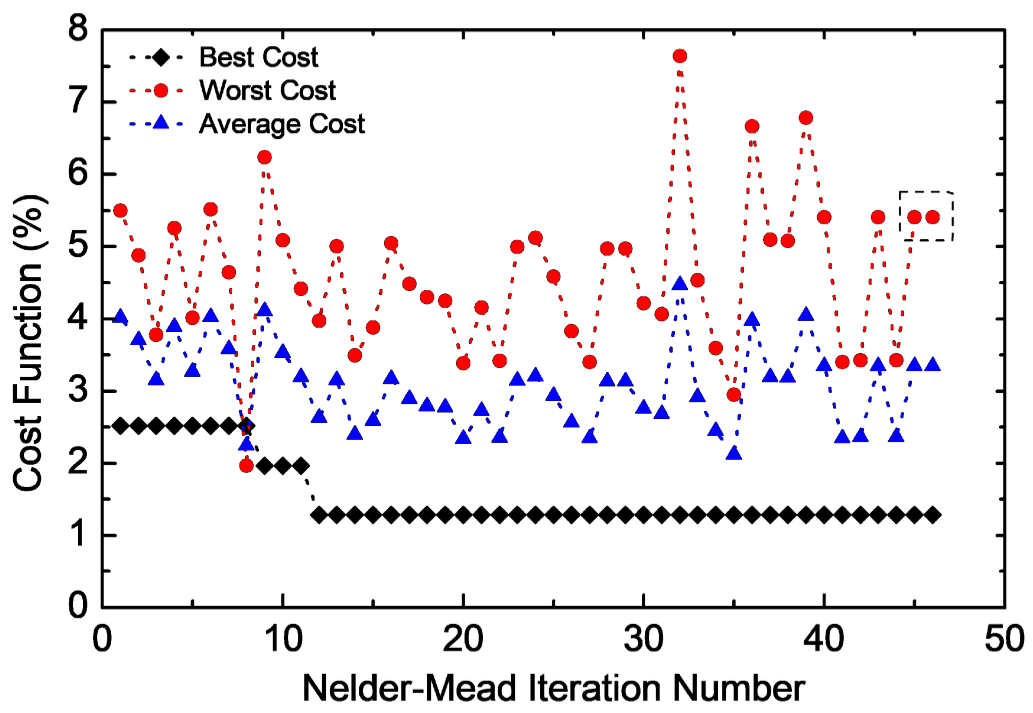}
  \caption{Cost function value variation while using Nelder-Mead algorithm}
  \label{fig4}
 \end{center}
\end{figure}

\subsubsection{Nelder-Mead optimization}

Genetic algorithm is a metaheuristic and may not converge to its local minimum always. As this was true in our case, we decided to further optimize using downhill simplex method also know as Nelder-Mead algorithm \cite{nelder1965simplex}. There exists many variations of Nelder-Mead, of which we used the one implemented \cite{lagarias1998convergence} in MATLAB software, and hence the variables and equations closely follow the notations therein. The resulting convergence of the results are shown in the Fig. \ref{fig4}. The parameters corresponding to the best cost in genetic algorithm population is used to create initial simplex. Also, the genetic algorithm optimization leads to a $\lambda$ value of 0.995393. Hence we choose $\lambda=1$ and avoided from further optimization. This also calls for the equivalence of both Eq. \ref{equation1} and Eq. \ref{equation5}.  For individual steps like reflection, expansion, contraction and shrink, CGMD simulations are performed and cost values are estimated. The trend of best, worst and average cost values during the Nelder-Mead optimization is shown in the \ref{fig4}. The iteration is stopped when worst cost converges with a tolerance value of 0.001. After 46 iterations, the Nelder-Mead converged with a best cost value of 1.287 \%. The corresponding optimized parameters of the MARTINI-E model is given in Table \ref{table1}.

\begin{table}[H]
    \centering
    \caption{Parameters of the MARTINI and MARTINI-E models}
    \begin{tabular}{c|c|c|c|c}
         Model & $\epsilon$ (kcal/mol) & $\sigma$ (A) & $\lambda$ & $\gamma$ ($kcal.fs.mol^{-1}.A^{-2}$) \\
         \hline
        MARTINI & 1.19503 & 4.7 & - & -\\
        MARTINI-E & 1.380404 & 4.815407 & 1.00 & 0.095097
        \end{tabular}
    
    \label{table1}
\end{table}

\section{Results}

After the parameterization of the model, the next step is comparing its performance with other CGMD models like original MARTINI model, mW and ML-mW models. For this purpose, a bulk liquid system, a film system and a vapo system is considered with system dimensions as shown in Table \ref{table2}. By varying the temperature from 350 K to 410 K, density, surface tension and enthalpy of vaporization is estimated. Note that the box dimensions of liquid system will change upon application of Nose-Hoover barostat.

\begin{table}[H]
    \centering
    \caption{Properties of various CGMD models used for simulation}
    \begin{tabular}{c|c|c|c|c|c|c}
    
         Model & $N_{vapor}$  & $N_{liquid}$  & $N_{film}$ & cubic $BOX_{vapor}$ & cubic $BOX_{liquid}$ & $BOX_{film}$  \\
         \hline
        
        MARTINI & $2000$ & $2000$ & $1045$ & $740.03\AA$ & $62.96\AA$ & $50\AA \times 50\AA \times 150\AA$\\
        
        MARTINI-E & $20$ & $2000$ & $1045$ & $200\AA$ & $73.452‬\AA$ & $50\AA \times 50\AA \times 200\AA$\\
        
        mW & $4005$ & $4005$ & $4005$ & $587.63\AA$ & $50\AA$ & $50\AA \times 50\AA \times 150\AA$\\
        
        ML-mW & $4005$ & $4005$ & $4005$ & $587.63\AA$ & $50\AA$ & $50\AA \times 50\AA \times 150\AA$
        \end{tabular}
    
    \label{table2}
\end{table}

\begin{figure}[H]
\begin{center}
\includegraphics[width=1\textwidth]{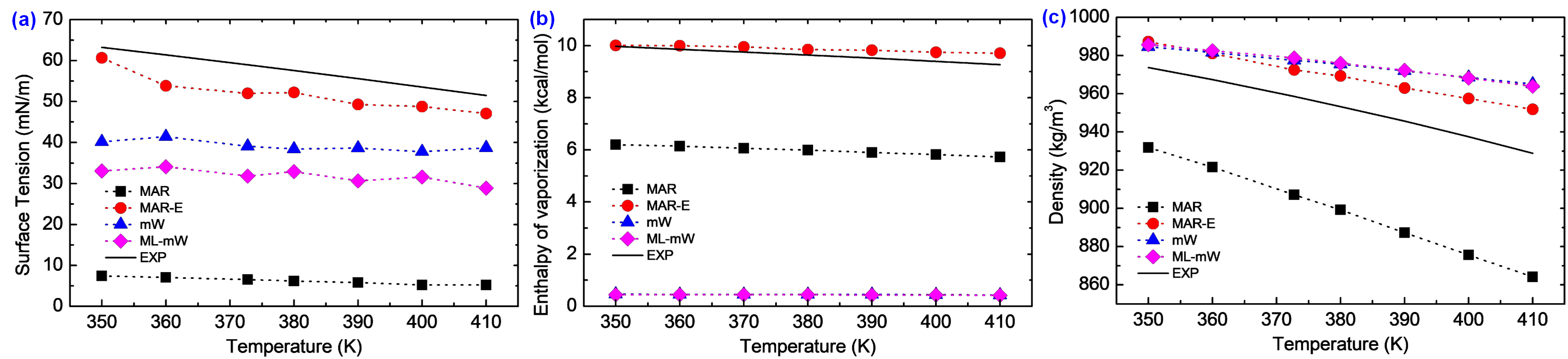}
  \caption{Variation of (a) surface tension, (b) enthalpy of vaporization and (c) density with temperature for various CGMD models}
  \label{fig5}
  \end{center}
\end{figure}

The results of surface tension, enthalpy of vaporization and density variation are shown in the Fig. \ref{fig5}. The legends MAR, MAR-E, mW, ML-mW and EXP corresponds to original MARTINI, MARTINI-E, mW, ML-mW and Experimental values respectively. The experimental values are shown as a bold solid line. At these temperature ranges, the mW and ML-mW performs poorly and MARTINI-E turns out to be the best among all for enthalpy of vaporization Fig. \ref{fig5}b. 

The surface tension results shows again that MARTINI-E model is closer to experimental results and is superior in comparison with other models Fig. \ref{fig5}. The density simulated by MARTINI-E is slightly above the experimental values as shown in the Fig. \ref{fig5}c, but still within 2\% error. These results show that the MARTINI-E model can be used for studies where enthalpy of vaporization, surface tension and density (compressibility) is important. Also, MARTINI-E is much accurate than other CGMD models for heat transfer.

\subsection{Sensitivity due to cutoff radius}

In the beginning, a cutoff radius of $12 \AA$ was used and later increased to $18 \AA$ for better results. To study the sensitivity of the cutoff radius, the liquid system (Fig. \ref{fig1}d) is simulated using NVT ensemble for 100,000 steps of equilibration and production.

\begin{figure}[H]
\includegraphics[width=\textwidth]{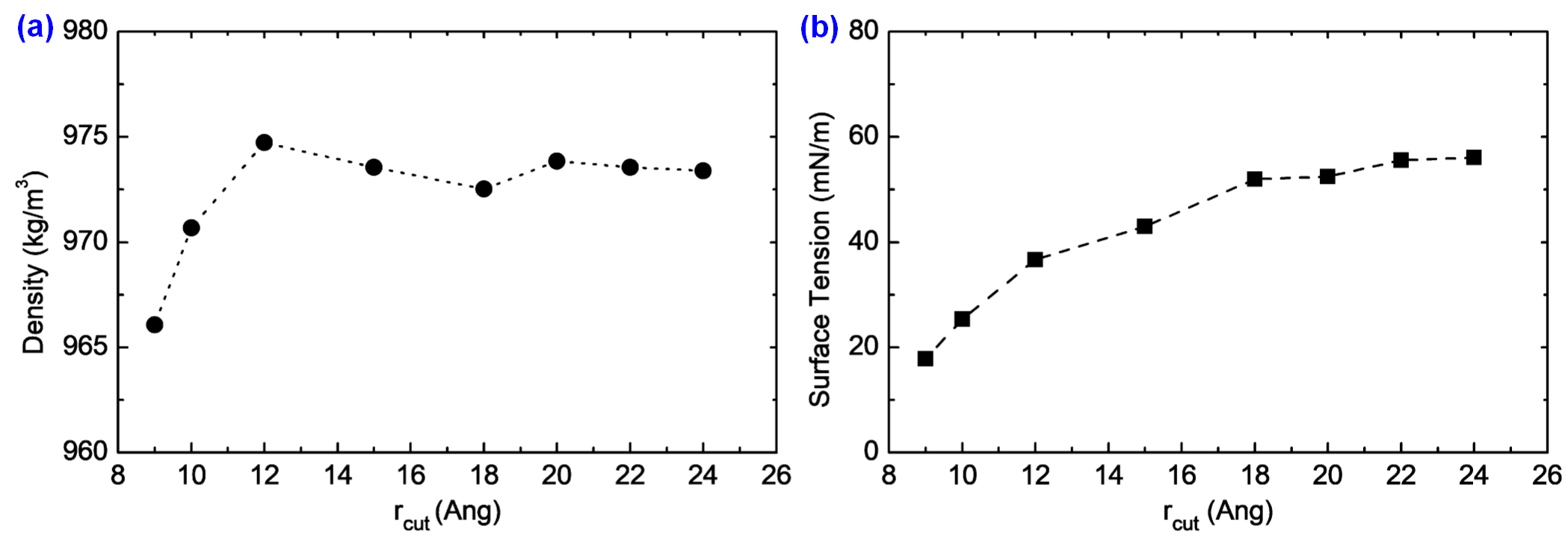}
  \caption{Sensitivity of (a) Density and (b) Surface tension with cut off radius}
  \label{fig6}
\end{figure}

The results of cutoff sensitivity study is shown in Fig. \ref{fig6}. The density shows variation until $18 \AA$ and surface tension also shows the same trend. Hence a minimum cutoff radius of $18 \AA$ is suggested for any CGMD simulations with MARTINI-E model.

\subsection{Sensitivity due to time step}

\begin{figure}[H]
\includegraphics[width=\textwidth]{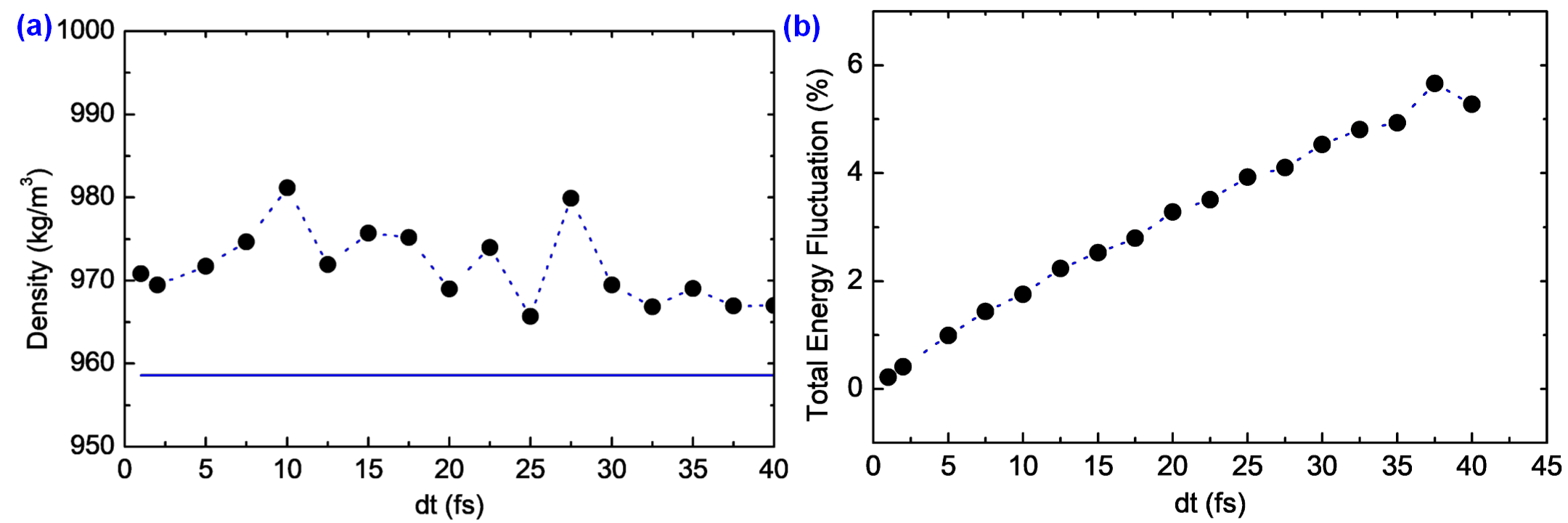}
  \caption{(a) Sensitivity of density and (b) Fluctuation (Coefficient of Variation) in total energy with respect to varying time step of integration}
  \label{fig7}
\end{figure}

The time step of integration for classical MARTINI models in the literature is suggested to be in the range of 20 fs to 30 fs \cite{winger2009using, marrink2010comment}. Here, a sensitivity study for the time step of integration for the MARTINI-E model is conducted. To unveil the true behavior of the system under different time steps, $NVE$ (microcanonical ensemble) instead of $NVT$ (canonical ensemble). The liquid system (Fig. \ref{fig1}d) is equilibrated using $NVT$ ensemble for 50,000 steps and followed by a production run of 50,000 steps in $NVE$ ensemble. The Coefficient of Variation of total energy, variation of density is estimated and shown in Fig. \ref{fig7}. The Coefficient of Variation (CV) of energy is estimated using below equation.

\begin{equation}
    E_{CV} = 100 \frac{\sigma_E}{\mu_E}
    \label{equation8}
\end{equation}

 Our results show that the coefficient of variation of the total energy of the system changes linearly with time step of integration. Also, most importantly, the system crashes due to instability after a time step of $40 fs$. Due to these reasons, eventhough the CV of the total energy is less than 5\% until $35 fs$, we recommend using time steps below $25 fs$ for the heat transfer simulations with MARTINI-E model.

\section{Validation with a lamellar system}

\begin{figure}[H]
\includegraphics[width=\textwidth]{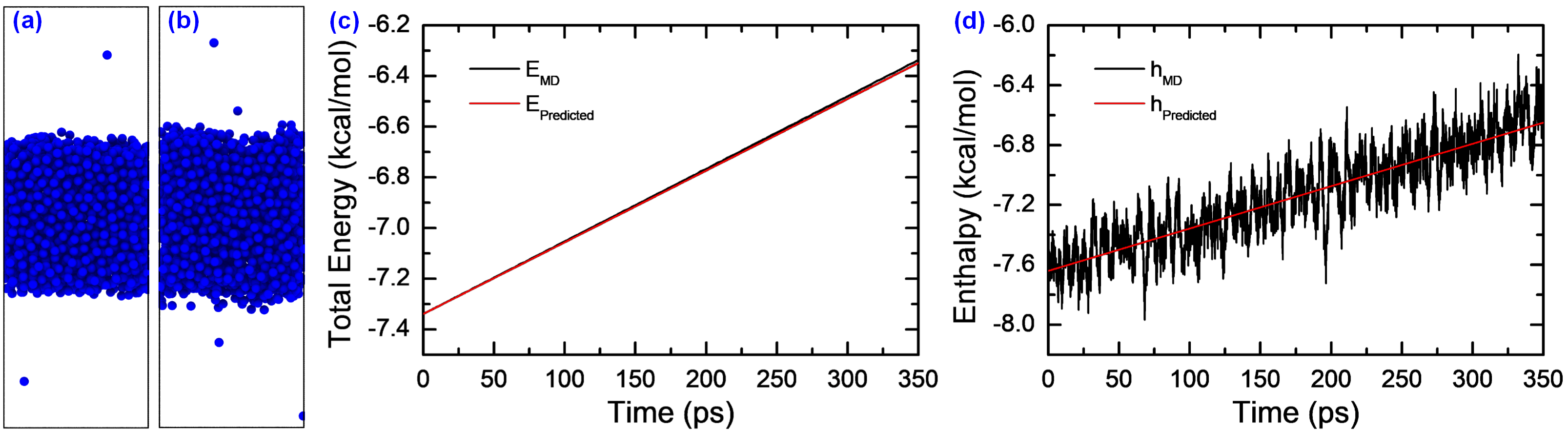}
  \caption{CGMD model of film (a) before and (b) after heat transfer. Evolution of (c) total energy with time and (d) enthalpy with time.}
  \label{fig8}
\end{figure}

This section explains the validation of the new MARTINI-E model based on basics of energy balance. A CGMD model as shown in Fig. \ref{fig8}a with dimensions of $62.68 \AA \times 62.68 \AA \times 182.68 \AA$ and liquid film thickness of $62.68 \AA$ at the center with 2000 beads. This system is equilibrated for 500,000 steps at 373.15 K using a time step of 10 fs. This equilibrated system is then supplied with energy at a constant rate using a heat exchange algorithm \cite{ikeshoji1994non}. According to energy balance of a closed system with no heat or work exchange with surroundings, the net heat input should be equal to net change in enthalpy.

As a test case, $1 kcal/mol$ of energy is supplied to the system in $353.6 ps$ representing a heat flux of $100 kW/cm^2$ based on the film's cross sectional base area. Figure \ref{fig8}b shows the system snapshot towards the end of the simulation. The time evolution of both total energy and enthalpy ($h = u + pv$) is shown in comparison with predicted rate (in red color) of increase in Fig. \ref{fig8}c,d. The results show that there is no deviation of enthalpy from the predicted rate of increase and is closely following the trend. The fluctuations seen in the Fig. \ref{fig8}d is due to the pressure, but the average value follows the trend. This study shows that the MARTINI-E model can conserve energy over long simulations, and most importantly the latent heat supplied is reflected as the change in enthalpy, which crucial in simulating evaporation processes.

\section{Discussion}

A simple extension of the original MARTINI model is developed by changing sigma and epsilon of LJ potential and adding a viscous force term to the force field. This has enabled to simulate surface tension, enthalpy of vaporization and density related to water at temperatures near 100 \degree C. The use of overly simplistic LJ potential limits the ability to match a larger number of parameters, say for example diffusion coefficient. This can be improvised using potential functions with more variables like Tersoff \cite{tersoff1988new} or Brenner \cite{brenner2002second} potential, however with an expense in computational cost. Though the current study doesn`t focus on dielectric properties, dipole moments, radial distribution function \cite{hockney1988computer}, etc. it may be beneficial to match them for a more accurate version of the CGMD model.

\section*{Conclusion}

The classical MARTINI model is successfully extended to use in a higher temperature range which will enable it to use for heat transfer simulations. Using a combination of genetic algorithm, artificial neural network and Nelder-Mead optimization, the original model is extended to a new model called MARTINI-E. This is validated against thermodynamic parameters including the co-existing points of temperature-pressure-density phase equilibrium, surface tension and enthalpy of vaporization. The model is validated further using the energy conservation and enthalpy change due to latent heat in a lamellar system. The new model is computationally faster like classical MARTINI model and at the same time can simulate accurately liquid-vapor co-existence and its interface.  This model can be used for large scale simulations to study the heat transfer characteristics at nanoscale and evaporation of thin films which are crucial for the heat transfer research.

\bibliographystyle{unsrt}

\bibliography{references}

\end{document}